\documentclass[epj]{svjour}
\usepackage{graphics}
\begin{document}
\title{Parity violation and the nature of charges}
\author{B. Desplanques
}                     
\institute{Laboratoire de Physique Subatomique et de Cosmologie 
(UMR CNRS/IN2P3--UJF--INPG),  \\
 F-38026 Grenoble Cedex, France}
\date{Received: date / Revised version: date}
%
\abstract{
The origin of parity violation in physics is still unknown. 
At the present time, it is introduced in the theory by requiring that 
the $SU(2)$ subgroup entering the description of interactions involves 
the left components. In the present contribution, one elaborates upon 
a suggestion made by Landau that particles and antiparticles could be 
like ``stereo-isomeric" molecules, which would naturally provides 
parity violation. Particles and antiparticles could thus be combinations 
of the parity doublets associated with a chiral symmetry realized 
in the  Wigner-Weyl mode. Consequences of such a description and the possible
problems it could raise are examined.
} 
\maketitle
%
\section{Introduction}
Thinking about the $\theta-\tau$ puzzle, Lee and Yang were led to suggest, 
50 years ago, that the parity symmetry could be violated in weak interactions 
\cite{LEEY}. This was confirmed a few months later by Wu {\it et al.} 
who observed a preferential direction for the $\beta$ emission in the decay of 
oriented $^{60}$Co \cite{WU}. Since then, the knowledge of the weak interaction 
has considerably increased. Together with strong and electromagnetic
interactions, it is embedded in the standard model, which is based on the 
$ SU(3)_{c} \otimes  SU(2)_{L} \otimes U(1)_{Y} $ gauge group with spontaneous 
symmetry breaking. Parity violation is ascribed to the appearance of the left
components associated with the $SU(2)$ weak isospin subgroup ($SU(2)_{L}$). 
It is sometimes considered that this property could be a low energy one 
and that the standard model should be completed by another  $SU(2)$ subgroup, 
invol\-ving this time the right components ($SU(2)_{R}$) \cite{LRSYM}. 
Implying further higher mass gauge bosons, the parity symmetry could thus 
be restored in high-energy processes. 

When trying to answer some question, it is not rare that one has
better to enlarge the problem. Another question, which is also a fundamental
one, is the nature of charges or, for our concern here, what makes a particle
different from its antiparticle. At present, particles and antiparticles appear
as solutions of the same equation, the Dirac one for instance, and they are
essentially cha\-racterized by some numbers,  reflecting their ``charges". 

Could it be that parity violation and the nature of charges be related to each
other? A hint is provided by the approximate conservation of the $PC$ symmetry, 
which suggests that the two operations $P$ and $C$ have a deep relationship. 
Which one however?

Some ideas along the above lines have been briefly proposed by Landau 
\cite{LAND} soon after the suggestion of the parity violation by Lee and Yang. 
He was considering that the difference between particles and 
antiparticles, as far as the space symmetry is concerned, ``is no greater 
than that due to chemical stereo-isomerism". Their ``charges" of
interest here are  especially the leptonic and baryonic ones.

In the present paper, we want to elaborate upon these ideas. In sect. 2, 
we make a couple of observations that relativize to some extent 
the origin of parity violation in physics. Section 3 is devoted to
describing a framework accounting for particles and antiparticles. 
The fourth section is concerned with the origin of parity violation 
in this framework and the emergence of the charge conjugation as 
a pa\-ri\-ty operation on an internal structure with a chiral character. 
Some results from a toy model are presented in sect. 5. 
An outlook is given in sect. 6.

\section{Some observations}
When asked about the origin of parity violation in physics, an answer 
often given is that nature is essentially left-handed. This is built in the
theory which relies on the gauge subgroup $SU(2)_{L}$. Had the theory been
written in terms of the charge conjugate fields ($\psi$ describes the creation 
of a particle and the destruction of the antiparticle, instead of the inverse),
the gauge subgroup would be  $SU(2)_{R}$. The left-handedness would have 
transformed into a right-handedness. The handedness thus depends on  what we
refer to as particles or antiparticles. Conside\-ring both particles and
antiparticles on the same footing, there would not be any preferred handedness.
Actually, this symmetry is nothing but the $PC$ symmetry which we assume to hold
here. It suggests that the charge conjugation operation, $C$, could have 
something to do with a usual parity operation.

Another observation concerns the mathematical definition of a parity operation.
As noticed by Lee and Wick \cite{LEEW}, if $P$ is a parity operation, the
product of $P$ by any unitary operator $U$ that leaves the interaction 
invariant is also a parity operation. This can be applied to the case where one 
takes for $U$ the charge conjugation operator $C$, which leaves invariant the
strong and electromagnetic interactions. Thus, the product $PC$ could be
considered as a parity operation. This definition is quite advantageous 
as the corresponding symmetry essentially  holds for all interactions. 
Again, this feature suggests that the charge conjugation could have something 
to do with a usual pa\-ri\-ty operation.
 
Following Landau and anticipating on next sections, we will speak 
of the product $PC$ as a combined parity operation.  We will denote it as 
$ PC=\mathcal{P}$, to remind that the operations  $P$  and  $C$ 
originate from a unique parity operation, acting however on different 
degrees of freedom.

\section{Framework for particles and antiparticles}

Models with chiral symmetry and parity symmetry have received a lot of 
attention in the strong interaction domain (QCD). In the simplest case 
and quite generally,  they are characterized by a conserved current, 
$J^{A}_{\mu}(x)$, and an anticommutation relation between the axial 
charge $Q^{A}= \int d \vec{x} \, J^{A}_0(x)$, and the (genuine) 
parity operator, $\mathcal{P}$:
\begin{equation}
[{\cal P},Q^A]_{+}= {\cal P} \; Q^A + Q^A \; {\cal P}=0. 
\end{equation}
A first realization of the symmetry, which stems from the above equation 
in the case $Q^A|{\rm state}\neq 0> \;\neq 0$, assumes that particles appear 
in the form of doublets: two equal-mass 
states of opposite parity or, equivalently, two states with opposite 
chiral charge, transforming into each other by a parity operation 
(Wigner-Weyl mode). It supposes $Q^A|0> \;= 0$. A second realization 
(Nambu-Jona-Lasinio mode) supposes  $Q^A|0> \;\neq 0$ and implies 
the existence of a Goldstone boson.  In absence of parity doublets,
this is the realization that has been retained in the QCD case 
where the Goldstone boson is  known to be the pion. 
While looking for parity doublets, it has been assumed that
the charges of these particles should be the same. Little attention 
has been given to the fact that they could be different and 
have a chiral character. Thus, particles and antiparticles could be 
associated to the states with opposite chiral charges of
an interaction exhibiting chiral symmetry or as combinations 
of the parity doublets it implies. 
It is evident that this chiral symmetry has nothing to do with the QCD one but
the existence of this one makes plausible the existence of an other 
one in a different sector of the interaction.

\section{Origin of parity violation and $C$ operation}
In atomic or nuclear physics, most of the observed parity-violating effects 
are ascribed to some parity admixture in the systems one is dealing with. The
question arises whether it could also be so at the more fundamental level.
In the present case, an important observation about pa\-ri\-ty conservation 
stems from eq. (1). The parity operation $\mathcal{P}$ and the chiral 
charge $Q^A$ do not commute; one cannot therefore generally have eigenstates 
of the parity and the chiral charge simultaneously. As most experiments 
involve ``charged" particles in the initial or final states, one should expect
to see some apparent parity violation, though pa\-ri\-ty is conserved at the
interaction level. The observed  pa\-ri\-ty violation would reflect the fact 
that these particles are not, intrinsically, eigenstates of the parity. 

While parity violation could be traced back to charges with a chiral nature, 
the problem would rather be to explain why it is so strong in some
cases (100\% in weak interactions) and absent in other cases (strong and
electroma\-gne\-tic interactions). An answer would require to consider 
some dynamics, perhaps unknown at the present time. A qualitative 
understanding can nevertheless be proposed. It supposes that the spin and
momentum properties of the chiral substructure decouple of those 
of the particle as a whole. In this decoupling limit, it is conceivable 
that the genuine parity  operation $\mathcal{P}$ factorizes into a part 
involving the external degrees of freedom (the usual $P$ operation) and a part 
involving the internal degrees of freedom (the usual $C$ operation). 
Of course, the decoupling cannot be complete and some parity violation 
reflecting the chiral nature of ``charges" could show up. 
It is noticed that some experiments do not necessarily involve ``charged" 
particles, like the neutral kaon decay into two mesons (the $\pi^+\,\pi^-$ 
state is globally considered here as a neutral one). Parity violation 
in this case could be observed too. Actually, this would be 
a true parity violation, which is known as a $PC$ violation \cite{PC}. 

The above explanation ascribing parity violation to an intrinsic chiral 
structure goes beyond Landau's proposal, which was likely to be only 
suggestive. Relying on some underlying chiral symmetry like
here makes it more systematic, as implied by observation and accounted for 
by the standard model. 

Developments presented in this section have a rather general character. 
They do not however provide any rea\-listic description of the physical world.
Actually, without entering much into details, many questions may be raised.
The first one concerns the nature of the chiral structure that could 
underly some charges and their quantization (possibly approximate).
A second non-trivial question has to do with the conservation of the axial
current that stems from the chiral symmetry we assumed, taking into account 
that particles have a mass. Other questions involve the dynamics. 
How this one makes the spin and momentum 
properties of the underlying chiral structure to approximately decouple 
from those of the particle as a whole? or also how particles get mass 
taking into account that the above chiral symmetry could be realized 
by a  pair of massless particles of opposite helicities (like for 
a Majorana neutrino). A further but different-type question concerns 
the equation that could play here the role of the Dirac equation. In the
picture developped above, particles and antiparticles are somewhat 
on the same footing, a unique combined parity operation, $\mathcal{P}$ 
($=PC$), transforming one into the other. In the context of the Dirac 
equation, antiparticles appear as holes in a Fermi sea filled up 
by negative-energy states.

\section{A toy model}
In order to get some insight on the answers to questions raised 
in the previous section, but also to emphasize possible problems, 
we here consider a particular model that is inspired from the 
Nambu-Jona-Lasinio model \cite{NAMB}, but with an essential 
difference \cite{SYMM}. As we want to associate ``charges"  
to some chiral structure, we change the role of the spin, whose projection
along some direction  was assumed to be conserved by elementary excitations 
in this model, with the  helicity one. 

In a mean-field approximation, it is conceivable that the resulting 
Lagrangian density reads:
\begin{equation}
{\cal L}(x)= \frac{1}{2} \left ( i \, \bar{\psi}(x) \gamma^{\mu} 
\partial_{\mu} 
\psi(x)
-m \; \bar{\psi}(x) \gamma . \epsilon \; \gamma_5 \psi(x)  \right)\,,
\end{equation}
where $\epsilon^{\mu} $ (with $\epsilon^2=-1$), which represents 
the polarization carried 
by the underlying chiral structure, could play a role similar 
to the deformation one in the case of intrinsically non-spherical nuclei. 
It is easily checked that the above Lagrangian exhibits some chiral symmetry 
and that the associated axial current and chiral charge (the helicity number 
in the present case) read:
\begin{equation} 
 J^{A}_{\mu}(x)= \frac{1}{2} \, \bar{\psi}(x)\gamma_{\mu} \gamma_5 \psi(x),  
\hspace{0.5cm}Q^{A}= \int d \vec{x} \, J^{A}_0(x).
\end{equation}  
In momentum space, it leads to the equation:
\begin{equation}
\left( \gamma . p-m \, \gamma .\epsilon \gamma_5 \right) 
\psi_{\epsilon}(p)=0.
\end{equation}
Solutions to this last equation are obtained under the condition 
$\epsilon \cdot p =0$, which corresponds to a decoupling 
between the spin carried by the underlying chiral structure 
and the momentum of the system as a whole. They can be expressed 
in terms of the standard solutions of the Dirac equation:
\begin{eqnarray}
\psi_{\epsilon}(p) \propto (1\!+\!\gamma_5) \, (1\!-\! \gamma .\epsilon) \, u(p)\, ,
\nonumber \\
\psi_{\epsilon}(p) \propto (1\!-\!\gamma_5) \, (1\!-\! \gamma .\epsilon) \, v(p)\, .\,
\end{eqnarray}
The appearance of the solutions of the Dirac equation for particles and
antiparticles is important as it allows one to make some relationship 
with its achievements. The pre\-sence of the front factors, 
$(1\!\pm\!\gamma_5)$, is essential with res\-pect to the present developments. 
It shows that parity violation could be built in the description 
of the particles themselves. The standard model of electro-weak interactions 
would thus be the effective one accounting for the above feature.

The toy model considered in this section casts also some light on the currents
and the appearance of the charge conjugation as an internal parity operation.
Considering matrix elements of the axial current with solutions for particles
and antiparticles, eq. (5), and integra\-ting over $\epsilon^{\mu} $ together 
with the further assumption 
$\epsilon \cdot w=0$ (decoupling of the spin carried by the chiral structure 
and the spin of the particle as a whole, $w$), one gets for a particle and its
antiparticle:
\begin{eqnarray}
 <S\!=\!\frac{1}{2},\,p\; |J^{A}_{\mu}(0)| \, S\!=\!\frac{1}{2},\,p> \;\; = 
\;\;\;\;\bar{u}(p)\gamma_{\mu}u(p)\, , \nonumber \\
\;\;\;\;\;\; {\rm and} \;\;\;\; -\bar{v}(p)\gamma_{\mu}v(p) \, .
\end{eqnarray}
The apparent change in the parity between the l.h.s. and r.h.s. could be
disturbing but it is noticed that the current at  the r.h.s. changes sign 
under a charge operation so that the behavior of both sides under 
the combined parity operation, $\mathcal{P}=PC$,   is the same. 
This is a consequence of a charge conjugation operation generated 
dynamically as a parity operation on an internal
chiral substructure. On the other hand, the appearance of a
vector current at the r.h.s. solves the major problem of the current
conservation that an axial current would raise unavoidably.

\section{Conclusion and outlook}
We presented a scheme where particles and antiparticles could be associated
to the parity partners of a chiral symmetry realized in the Wigner-Weyl mode. 
This provides a natural explanation for parity violation in physics. 
We describe how charge conjugation, which is no more than a parity operation 
on a substructure with a chiral cha\-racter, emerges in this picture. 
Some of the questions and problems that could be raised have been outlined.

There is far from the present sketch to a realistic theory and it is not 
clear whether one can deal with some constraints on combining space-time 
and internal symmetries \cite{COLE}. We nevertheless believe that some 
of the ideas considered here could cast some new light in developping theories. 
As examples, we notice that there is no point to restore the usual parity 
symmetry, its violation being intrinsically tied to the chiral substructure 
underlying some ``charges". There is no need for introducing right-handed
currents in the simplest case (see remark on $SU(2)_{R}$), and the difference
between $V$ and $A$ currents could vanish at some point. Despite their
speculative character, we guess it was appropriate to remind these ideas 
in this year where one celebrates 50 years of parity violation.


\end{document}